\documentclass[aps,prb,twocolumn,superscriptaddress,showpacs,floatfix,longbibliography]{revtex4-2}
\usepackage{amsmath,amssymb,amsfonts,float,graphics,epsfig,epstopdf,color,verbatim,tabularx,bm,multirow,appendix,hyperref}
\usepackage[normalem]{ulem}
\usepackage{lmodern}
\usepackage{ytableau}
\usepackage{color}
\usepackage{bm}
\usepackage{wasysym}
\usepackage{mathrsfs}

\newcommand{\beq} {\begin{equation}}
	\newcommand{\eeq} {\end{equation}}
\newcommand{\bea} {\begin{eqnarray}}
	\newcommand{\eea} {\end{eqnarray}}
\newcommand{\be} {\begin{equation}}
	\newcommand{\ee} {\end{equation}}

\DeclareMathOperator{\Tr}{Tr}

\begin{document}
	
\title {Delay Update in Determinant Quantum Monte Carlo}
\author{Fanjie Sun}
\affiliation{Key Laboratory of Artificial Structures and Quantum Control (Ministry of Education),
	School of Physics and Astronomy, Shanghai Jiao Tong University, Shanghai 200240, China}
	\author{Xiao Yan Xu}
	\email{xiaoyanxu@sjtu.edu.cn}
\affiliation{Key Laboratory of Artificial Structures and Quantum Control (Ministry of Education),
	School of Physics and Astronomy, Shanghai Jiao Tong University, Shanghai 200240, China}
\affiliation{Hefei National Laboratory, Hefei 230088, China}

\date{\today}
		
\begin{abstract}
Determinant quantum Monte Carlo (DQMC) is a widely used unbiased numerical method for simulating strongly correlated electron systems. However, the update process in DQMC is often a bottleneck for its efficiency. To address this issue, we propose a generalized delay update scheme that can handle both onsite and extended interactions. Our delay update scheme can be implemented in both zero-temperature and finite-temperature versions of DQMC. We apply the delay update scheme to various strongly correlated electron models and evaluate its efficiency under different conditions. Our results demonstrate that the proposed delay update scheme significantly improves the efficiency of DQMC simulations, enabling it to simulate larger system sizes.
\end{abstract}

\maketitle

\section{Introduction}
Strongly correlated electron systems have garnered significant attention over the past few decades. These systems exhibit a rich variety of phases, such as high-temperature superconductivity~\cite{Bednorz1986BaLaCuO,Bednorz1988highTc,johnston2010puzzle,Lee2006doping,sun2023signatures}, fractional quantum Hall~\cite{tsui82,Moore1991,Stormer1999,Hansson2017}, quantum spin liquids~\cite{Anderson1973,Kitaev2006anyons,Balents2010,zhou2017quantum,broholm2020quantum,Savary2017,broholm2020quantum} and so on. Their study has led to a deeper understanding of the underlying profound physics governing their behavior. Due to the non-perturbative nature of strongly correlated electron systems, the development of advanced and unbiased simulation methods is required to accurately study their properties.

Determinant Quantum Monte Carlo (DQMC)~\cite{Blankenbecler1981qmcc,scalapino1981monte,Hirsch1983HStransform,Sugiyama1986,Sorella1988numerical,Sorella1989a,loh1992stable,koonin1997shell,Assaad2008compumanybody}, often also referred to as auxiliary field quantum Monte Carlo (AFQMC), has emerged as a powerful and widely-used technique for simulating strongly correlated electrons. It provides crucial insights into the behavior of these systems. This method has been successfully applied to a variety of interesting models, enabling researchers to explore novel phases and phase transitions~\cite{Hirsch1985hubbard,white19892dhubbard,assaad1999quantum,Scalapino2007,zheng2011,Hohenadler2011,Berg2012signproblem,hohenadler2013correlation,Assaad2013,LeBlanc2015,li2015solving,wang2015split,he2016bona,Otsuka2016qc,assaad2016simple,Zhou2016mott,gazit2017emergent,berg2019monte,li2019sign,xu2019revealing,liu2019superconductivity,xu2019monte,chen2019charge,zhang2019charge,xu2021competing,zhang2021momentum,hofmann2022fermionic,mondaini2022quantum}. However, despite its many successes, DQMC is not without its limitations, even when focusing only on sign-problem-free models. Due to the inherently high computational complexity associated with this approach, simulations are typically restricted to relatively small system sizes. This limitation restricts our ability to accurately extrapolate to the thermodynamic limit, which is necessary to determine the properties of phases and phase transitions with greater accuracy.

The computational complexity of DQMC primarily arises from the need to update and manipulate large Green's function matrices during the simulation process. In DQMC, we use Trotter decomposition such that the inverse temperature $\beta$ is divided into many small slices (e.g. $\beta = L_\tau a_\tau$, where $a_\tau$ is a small imaginary time slice, $L_\tau$ is the total number of imaginary time slices), and Hubbard-Stratonovich (HS) transformation is used to decouple the interaction term by introducing auxiliary fields. We usually need $\mathcal{O}(\beta N)$ number of auxiliary fields, where $N$ represents the system size. After HS transformations, the fermion degrees of freedom can be traced out~\cite{Blankenbecler1981qmcc,scalapino1981monte,Hirsch1983HStransform,Sugiyama1986,Sorella1988numerical,Sorella1989a,loh1992stable,koonin1997shell,Assaad2008compumanybody}, and the partition function of the system can be written as a sum over configurations of auxiliary fields with the Boltzmann weight being the inverse of the determinant of the Green's function matrix. In DQMC, Metropolis sampling with local updates is typically employed, where auxiliary fields are proposed to update one by one. To calculate the acceptance ratio, one has to calculate the determinant ratio of Green's function matrices. Additionally, the calculation of physical quantities necessitates the use of Green's function matrices. Thus, it is crucial to keep track of Green's function matrices during the MC sampling. 

However, the direct calculation of the determinant ratios and the Green's function matrices has a very high computational complexity of $\mathcal{O}(N^3)$.  To reduce this computational complexity, a common practice is the so called fast update scheme. In this scheme, if the proposed update is accepted, by using the Woodbury matrix identity, the Green's function matrix before and after update only differs by a low-rank matrix with rank $k$ determined by the number of sites connected to each single auxiliary field. This structure allows for the calculation of acceptance ratios -- related to determinant ratios --  with lower complexity $\mathcal{O}(k^3)$ (see Eq.~\eqref{eq:ratio} for more details). As a result, the calculation of updated Green's function also has a reduced complexity of $\mathcal{O}(k N^2)$ due to this low-rank difference. Thus, we also refer to fast update as `low rank update'. It is worthwhile to noticing that for DQMC simulation of models with local interaction where $k\sim\mathcal{O}(1)$, one Monte Carlo step (which is defined as a whole sweep of all auxiliary fields back and forth, and we call that a whole sweep of all auxiliary fields back and forth as a sweep in this work.) has the computational complexity of $\mathcal{O}(\beta N^3)$, while for models with long range interactions~\cite{zhang2021momentum,hofmann2022fermionic},  where $k\sim\mathcal{O}(N)$, this complexity increases to $\mathcal{O}(\beta N^4)$. The latter situation is relatively rare and is not within the scope of our consideration in this work. 

In fact, there is no need to update the Green's function immediately after each acceptance. It was pointed out that the successive low-rank updates of Green's function after each local update can be delayed as the determinant ratio calculation only requires the diagonal part of the Green's function~\cite{alvarez2008new,nukala2009fast}. This technique is known as the delay update. In the delay update approach, after each local update, only the diagonal part of the Green's function is updated, along with recording some intermediate vectors that are useful for the final update of the entire Green's function. The delay update has been successfully implemented in Hirsch-Fye QMC~\cite{alvarez2008new,nukala2009fast} and continuous-time QMC~\cite{gull2011submatrix}, and it can significantly reduce the prefactor of total computational complexity. However, those implementations only consider finite temperature and onsite interactions with $k=1$. How to generalize this to multi-site extended interactions with $k>1$ and to zero-temperature scenarios remains an incomplete task. Since a delay update with $k>1$ or applied in a zero-temperature case has never been considered in previous literature, its performance remains an open question and constitutes another objective of this paper.

 In this work, we consider general $k$ delay update, such that it can apply to general strongly correlated electron models with extended interactions. We implement the general delay update in the framework of DQMC both in the finite and zero-temperature versions. The general delay update can also be applied straightforwardly to Hirsch-Fye QMC, continuous-time QMC, as well as other QMC methods that involve the fermion determinant.

The remainder of the article is organized as follows. In Sec.~\ref{sec:method}, we introduce the basic formalism of the delay update. In Sec.~\ref{sec:results}, we compare the efficiency of the delay update with the conventional fast update on both the Hubbard model and the spinless $t$-$V$ model on a two-dimensional square lattice. At last, we make a brief conclusion and discussion in Sec.~\ref{sec:conclusion}.

\section{Method}
\label{sec:method}
Before introducing the basic formalism of generalized delay update, we review the fast update first~\cite{Blankenbecler1981qmcc,scalapino1981monte,Hirsch1983HStransform,Sugiyama1986,Sorella1988numerical,Sorella1989a,loh1992stable,koonin1997shell,Assaad2008compumanybody}. Without loss of generality, we first consider the finite temperature version of DQMC (DQMC-finite-T), and discuss the zero-temperature case only when necessary. In DQMC, the interacting fermion problem is mapped to a non-interacting fermion problem coupled with auxiliary fields (${\bm{s}}$). Taking the Hubbard model~\cite{Gutzwiller1963effect,Hubbard1963electron,arovas2022hubbard,qin2022hubbard}  as an example, with Hamiltonian
\begin{equation}
H_{tU}=H_0+H_U
\end{equation}
where $H_0=-t\sum_{\langle i,j\rangle,\alpha}(c_{i,\alpha}^{\dagger}c_{j,\alpha}+\text{H.c.})$ and $H_U=\frac{U}{2}\sum_i(n_i-1)^2$. In the Hamiltonian, $c_{i,\alpha}^{\dagger}$ creates an electron on site $i$ with spin polarization $\alpha=\uparrow/\downarrow$, $n_i=\sum_\alpha c_{i\alpha}^\dagger c_{i\alpha}$ is the fermion number density operator, $t$ is the nearest-neighbour (NN) hopping term, and $U$ is the amplitude of onsite Hubbard repulsive interaction.
In this work, we use the following HS transformation
\begin{equation}
\begin{aligned}
&e^{-a_\tau\frac{U}{2}\sum_i(n_i-1)^2}\\
=&\frac{1}{4}\sum_{s=\pm1,\pm2}\gamma(s)e^{-\text{i}\alpha_U\sum_i \eta(s)(n_i-1)}+O(a_\tau^4)
\end{aligned}
\end{equation}
where $s$ represents an auxiliary field, $\alpha_U=\sqrt{a_\tau U}$, $\gamma(\pm1)=1+\frac{\sqrt{6}}{3}$, $\gamma(\pm2)=1-\frac{\sqrt{6}}{3}$, $\eta(\pm1)=\pm\sqrt{2(3-\sqrt{6})}$, $\eta(\pm2)=\pm\sqrt{2(3+\sqrt{6})}$.
After Trotter decomposition and HS transformation, the fermion degrees of freedom can be easily traced out, so the partition function becomes
\begin{equation}
Z=\sum_{\bm{s}}w_{b}[\bm{s}]w_{f}[\bm{s}]
\end{equation}
where we denote all auxiliary fields $\{\bm{s}_{i,l}\}$ at every site $i$ and time slices $l$ as $\bm{s}$, and define the scalar weight
\begin{equation}
w_b\left[\bm{s}\right]\equiv \prod_{i,l}(\gamma(\bm{s}_{i,l})e^{-i\alpha_U \eta(\bm{s}_{i,l})})
\end{equation} 
and the determinant weight 
\begin{equation}
\label{z1}
\begin{aligned}
w_f\left[\bm{s}\right]&= \Tr\left[D_{\bm{s}}(\beta,0)\right]\\
&=\det\left[I+B_{\bm{s}}(\beta,0)\right].
\end{aligned}
\end{equation}
$B_{\bm{s}}(\beta,0)$ is the time evolution matrix and $D_{\bm{s}}(\beta,0)$ is the time evolution operator, and the general time evolution matrix and operator  $B_{\bm{s}}(\tau_2,\tau_1)$,  $D_{\bm{s}}(\tau_2,\tau_1) $($\tau_2 \ge \tau_1$) is defined as 
\begin{equation}
B_{\bm{s}}(\tau_2,\tau_1) = B_{\bm{s}}^{l_2} B_{\bm{s}}^{l_2-1} \cdots B_{\bm{s}}^{l_1+1}
\end{equation} 
\begin{equation}
D_{\bm{s}}(\tau_2,\tau_1) = D_{\bm{s}}^{l_2} D_{\bm{s}}^{l_2-1} \cdots D_{\bm{s}}^{l_1+1}
\end{equation} 
where $\tau_2=l_2 a_\tau$, $\tau_1=l_1 a_\tau$, $B^{l}_{\bm{s}}= e^{\bm{V}({\bm{s}})}e^{-a_\tau K}$ and $D^{l}_{\bm{s}}= e^{\bm{c}^{\dagger}\bm{V}({\bm{s}})\bm{c}}e^{-a_\tau \bm{c}^{\dagger}K\bm{c}}$, where $K$ is the coefficient matrix for non-interacting part of the Hamiltonian $ H_0=\sum_{j,k}c_j^{\dagger}K_{j,k}c_k=\bm{c}^{\dagger}K\bm{c}$, and $\bm{V}({\bm{s}})$ is the coefficient matrix for interacting part after HS transformation, $\text{i}\alpha_U\sum_i \eta(\bm{s})n_i=\sum_{j,k}c_j^{\dagger}\bm{V}(\bm{s})_{j,k}c_k=\bm{c}^{\dagger}\bm{V}(\bm{s})\bm{c}$ ($\bm{V}(\bm{s})$ can become non-diagonal in multi-site extended interaction models such as the $t$-$V$ model, which can be seen in Eq.~\eqref{eq:tv}). Note that we have absorbed the spin index into the site index for convenience. Above formalism directly applies to general interacting models, so in the following, we will not limited to Hubbard model, but consider general interactions which may have extended interactions. 

Consider a local update of $\bm{s}$, which only changes a specific component $\bm{s}_{i,l}$. We denote the effect of this change on $e^{\bm{V}({\bm{s}}_{i,l})}$ by $\Delta(\bm{s},\bm{s}')=e^{\bm{V}({\bm{s}'})}e^{-\bm{V}({\bm{s}})}-I$, where $\bm{s}'$ represents the auxiliary fields after the update. Then the determinant ratio can be calculated as
\begin{equation}
\frac{w_f[\bm{s}']}{w_f[\bm{s}]}=\det\left[I+\Delta(\bm{s},\bm{s}')(I-G_{\bm{s}}(\tau,\tau))\right]
\end{equation}
where the equal-time Green's function $G_{\bm{s}}(\tau,\tau)$ is a matrix with the elements $G_{\bm{s},ij}(\tau,\tau)\equiv \langle c_i(\tau)c_j^{\dagger}(\tau)\rangle_{\bm{s}}$, where
\begin{equation}
\langle \bm{O}\rangle_{\bm{s}}\equiv \frac{\Tr[D_{\bm{s}}(\beta,\tau)\bm{O}D_{\bm{s}}(\tau,0)]}{\Tr[D_{\bm{s}}(\beta,0)]}
\end{equation}
and $G_{\bm{s}}(\tau,\tau)$ can be calculated as~\cite{Blankenbecler1981qmcc,scalapino1981monte,Hirsch1983HStransform,Sugiyama1986,Sorella1988numerical,Sorella1989a,loh1992stable,koonin1997shell,Assaad2008compumanybody} 
\begin{equation}
\label{eq:G}
G_{\bm{s}}(\tau,\tau) = \left(I + B_{\bm{s}}(\tau,0)B_{\bm{s}}(\beta,\tau) \right)^{-1}
\end{equation}

To keep the notations succinct, we will omit the explicit dependence on ${\bm{s}}$ and $\bm{s}'$ for $G_{\bm{s}}$ , $\Delta(\bm{s},\bm{s}')$, $B_{\bm{s}}$ and all other related matrices. We only add back the explicit dependence when necessary. We assume $\Delta$ is a sparse matrix with only $k$ non-zero elements, all of which are located on the diagonal for general cases including the Hubbard model. We can obtain this form usually for local interactions by unitary transformations. We denote the  positions on the lattice in real place of the $k$ non-zero elements as $x_1, x_2, \ldots, x_k$. Due to the sparse nature of $\Delta$, using the Sylvester's determinant theorem, the determinant ratio can be simplified to
\begin{equation}
\label{eq:ratio}
\frac{w_f[\bm{s}']}{w_f[\bm{s}]}=\det\left[S\right]
\end{equation}
where $S$ is only a $k$-dimensional matrix.
\begin{equation}
\label{eq:dupr}
S=I_{k\times k}+\mathcal{V}D.
\end{equation}
where $I_{k\times k}$ is the $k$-dimensional identity matrix, $\mathcal{V}$ and $D$ are $k$-dimensional matrices written as
\begin{equation}
\mathcal{V}=\left[\begin{array}{ccc}
-(G_{x_{1}x_{1}}-1) & -G_{x_{1}x_{2}} & \cdots\\
-G_{x_{2}x_{1}} & -(G_{x_{2}x_{2}}-1) & \cdots\\
\vdots & \vdots & \ddots
\end{array}\right]_{k\times k}
\end{equation}
\begin{equation}
\label{eq:dmat}
D=\left[\begin{array}{ccc}
\Delta_{x_{1}x_{1}}\\
 & \Delta_{x_{2}x_{2}}\\
 &  & \ddots
\end{array}\right]_{k\times k}
\end{equation}

If the proposed update is accepted, the fast update scheme employs the Woodbury matrix identity. Consequently, the Green's function after the update (denoted as $G'$) only differs from the Green's function before the update (denoted as $G$) by a rank-$k$ matrix. This matrix is the product of three matrices, $\mathbb{U}\mathbb{S}\mathbb{V}$.
\begin{equation}
\label{eq:dupg}
G' = G + \mathbb{U}\mathbb{S}\mathbb{V}
\end{equation}
where $\mathbb{U}$ is an $N \times k$ matrix, $\mathbb{S}$ is a $k\times k$ matrix, and $\mathbb{V}$ is a $k \times N$ matrix.
\begin{align}
\mathbb{U} & =[G_{:,x_{1}}|G_{:,x_{2}}|\cdots]_{N\times k}
\end{align}
\begin{equation}
\mathbb{S}= D S^{-1}
\end{equation}
\begin{equation}
\mathbb{V}=\left(\left[G_{x_{1},:}-e_{x_{1}}|G_{x_{2},:}-e_{x_{2}}|\cdots\right]^{T}\right)_{k\times N}
\end{equation}
where $e_{x_j}$ denotes a unit row vector with only one non-zero element at position $x_j$, $G_{:,x_j}$ denotes column $x_j$ of $G$, and $G_{x_j,:}$ denotes row $x_j$ of $G$. 

\begin{figure}[htp!]	\includegraphics[width=1.0\columnwidth]{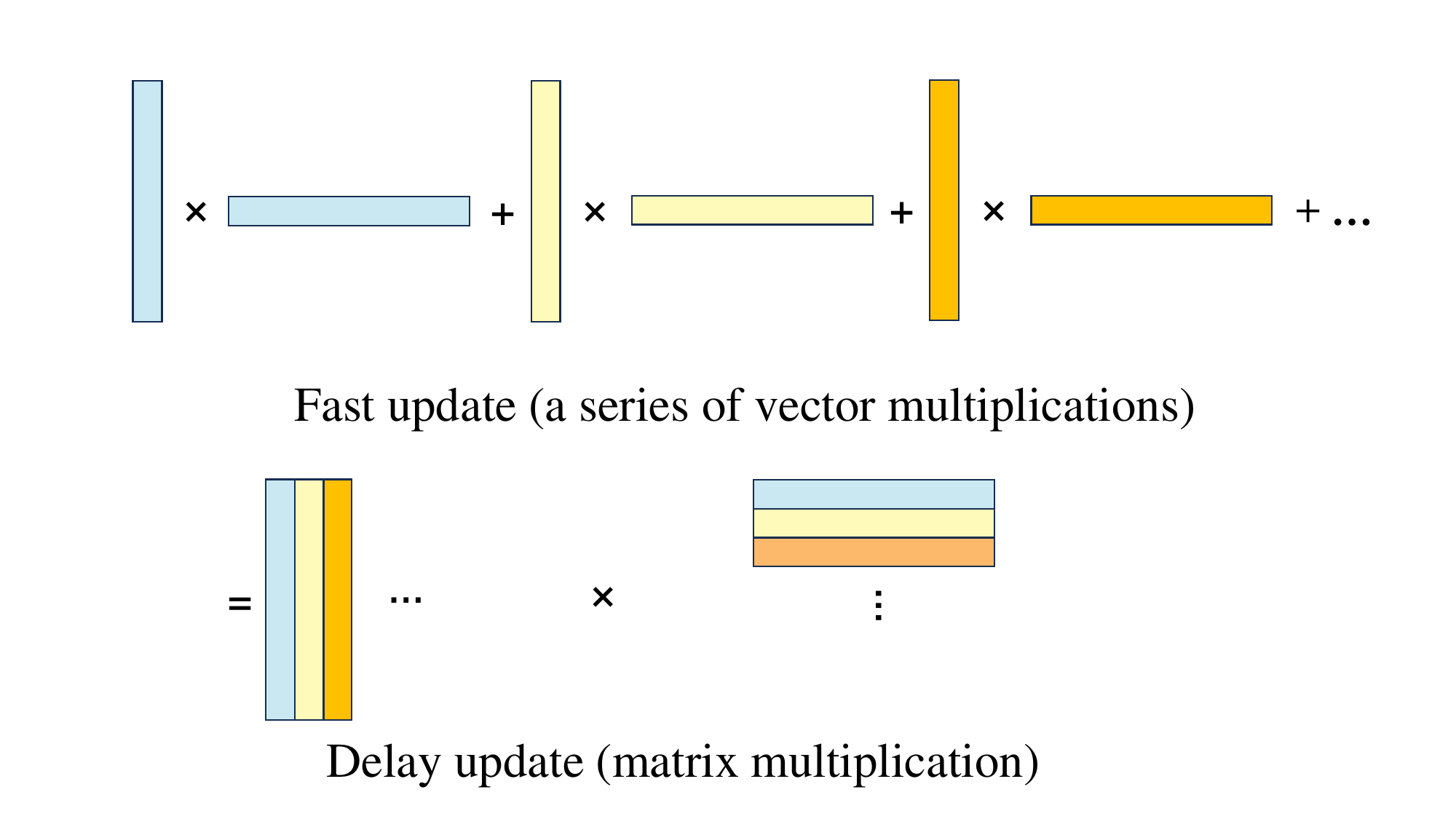}
	\caption{The schematic diagram of the delay update process. Compared to the fast update process, the delay update approach transforms a series of vector multiplications into matrix multiplications.}
	\label{fig:fig1}
\end{figure}

As we can see in Eqs.~(\ref{eq:ratio}-\ref{eq:dmat}), the determinant ratio only involves a very small part of the Green's function. Therefore, the delay update scheme applies. In delay update, we only update the entire Green's function occasionally. After each local update, we collect the matrices $\mathbb{U}$, $\mathbb{S}$, and $\mathbb{V}$. These are used to update the entire Green's function once a predetermined number of such matrices have been accumulated. We denote this predetermined number as $n_d$. 

Let us consider the delay update step by step. For convenience, we use the superscript $(i)$ to denote the relevant matrices generated in the $i$-th update. At the beginning, we have the initial $G^{(0)}$ and use it to calculate $S^{(1)}$, then the determinant ratio is $\det(S^{(1)})$. If the update is accepted, calculate $D^{(1)}$, $\mathbb{U}^{(1)}$, $\mathbb{S}^{(1)}$ and $\mathbb{V}^{(1)}$. Assume that $i-1$ updates, which is less than $n_d$, have already been accepted. Consequently, we have stored the sets $\{\mathbb{U}^{(1)},\ldots, \mathbb{U}^{(i-1)}\}$, $\{\mathbb{S}^{(1)},\ldots, \mathbb{S}^{(i-1)}\}$ and $\{\mathbb{V}^{(1)},\ldots, \mathbb{V}^{(i-1)}\}$. Now we consider the $i$-th step and present the following recurrence relations. At the $i$-th step, we calculate
\begin{equation}
S^{(i)}=I_{k\times k}+\mathcal{V}^{(i)}D^{(i)}
\end{equation}
with
\begin{equation}
\mathcal{V}^{(i)}=\left[\begin{array}{ccc}
-(G_{x_{1}^{(i)}x_{1}^{(i)}}^{(i-1)}-1) & -G_{x_{1}^{(i)}x_{2}^{(i)}}^{(i-1)} & \cdots\\
-G_{x_{2}^{(i)}x_{1}^{(i)}}^{(i-1)} & -(G_{x_{2}^{(i)}x_{2}^{(i)}}^{(i-1)}-1) & \cdots\\
\vdots & \vdots & \ddots
\end{array}\right]_{k\times k}
\end{equation}
\begin{equation}
D^{(i)}=\left[\begin{array}{ccc}
\Delta_{x_{1}^{(i)}x_{1}^{(i)}} & 0 & \cdots\\
0 & \Delta_{x_{2}^{(i)}x_{2}^{(i)}} & \cdots\\
\vdots & \vdots & \ddots
\end{array}\right]_{k\times k}
\end{equation}
The determinant ratio is $\det\left[S^{(i)}\right]$. If the update is accepted, we calculate
\begin{align}
\mathbb{U}^{(i)} & =[G_{:,x_{1}^{(i)}}^{(i-1)}|G_{:,x_{2}^{(i)}}^{(i-1)}|\cdots]_{N\times k}
\end{align}
\begin{equation}
\mathbb{S}^{(i)}=D^{(i)}\left(S^{(i)}\right)^{-1}
\end{equation}
\begin{equation}
\mathbb{V}^{(i)}=\left(\left[G_{x_{1}^{(i)},:}^{(i-1)}-e_{x_{1}^{(i)}}|G_{x_{2}^{(i)},:}^{(i-1)}-e_{x_{2}^{(i)}}|\cdots\right]^{T}\right)_{k\times N}
\end{equation}
where $e_{x_j^{(i)}}$ denotes a unit row vector with only one non-zero element at position $x_j^{(i)}$, $G_{:,x_j^{(i)}}^{(i-1)}$ denotes column $x_j^{(i)}$ of $G^{(i-1)}$, and $G_{x_j^{(i)},:}^{(i-1)}$ denotes row $x_j^{(i)}$ of $G^{(i-1)}$. Above, we calculate parts of the Green's function before the $i$-th step update using the following formulas,
\begin{equation}
G_{:,x_{j}^{(i)}}^{(i-1)}=G_{:,x_{j}^{(i)}}^{(0)}+\sum_{m=1}^{i-1}\mathbb{U}_{:,\mu}^{(m)}\mathbb{S}_{\mu\nu}^{(m)}\mathbb{V}_{\nu x_{j}^{(i)}}^{(m)}
\end{equation}
\begin{equation}
G_{x_{j}^{(i)},:}^{(i-1)}=G_{x_{j}^{(i)},:}^{(0)}+\sum_{m=1}^{i-1}\mathbb{U}_{x_{j}^{(i)}\mu}^{(m)}\mathbb{S}_{\mu\nu}^{(m)}\mathbb{V}_{\nu,:}^{(m)}
\end{equation}
where $j=1,2,\ldots,k$. After $n_d$ steps, we update the entire Green's function
\begin{align}
G^{(n_d)}=G^{(0)}+\sum_{m=1}^{n_d}\mathbb{U}^{(m)}\mathbb{S}^{(m)}\mathbb{V}^{(m)}
\end{align}
This can be done by matrix product, as shown in Fig.~\ref{fig:fig1}. 

Here we make a brief discussion about the computational complexity for delay update. The total overhead for preparing the intermediate matrices mainly comes from calculating part of the Green's function, which has the computational complexity $\mathcal{O}(n_d^2 N)$, and after $n_d$ steps, the calculation of the whole Green's function costs $\mathcal{O}(n_d N^2)$. By going through all auxiliary fields, we need to multiply the above complexity by a factor of $\mathcal{O}(\beta N/n_d)$, resulting in $\mathcal{O}(\beta n_d  N^2)$ for overhead and $\mathcal{O}(\beta  N^3)$ for Green's function updates.
In the DQMC, the delay update and fast update share the same computational complexity. However, in delay update a series of vector outer-products are transformed into matrix products as shown in Fig.~\ref{fig:fig1}, which significantly reduces the computation time. The main reason is that, for vector operations, Level 1 BLAS is used, while for matrix operations, Level 3 BLAS is used, which can optimize the usage of cache~\cite{van2008science}. 
Furthermore, we note that the optimized value of $n_d$ is limited by the size of the cache. Based on our test on the platforms with  roughly similar sizes of cache listed in Appendix~\ref{sec:app2}, setting $n_d$ to $2^\lambda$, where $\lambda=\min[6,[\log_2\frac{N}{20}]]$ , has proven to be a good choice. For example, for $N=42^2$, $n_d$ can be set to 64.

The above delay update scheme can also be applied to zero-temperature version of DQMC. For the zero-temperature version of DQMC (DQMC-zero-T), the normalization factor of ground state wavefunction $|\Psi_0\rangle$ plays a similar role to that of the partition function in the finite-temperature case.
\begin{equation}
\label{phi0}
\langle \Psi_0 | \Psi_0 \rangle = \sum_{\bm{s}} \det\left[ P^\dagger B_{\bm{s}}(2\Theta,0) P \right]
\end{equation} 
where $|\Psi_0\rangle=e^{-\Theta H}|\Psi_T\rangle$ is obtained by projection on a trial wavefunction $|\Psi_T\rangle=\prod_{i=1}^{N_\text{p}}({\bm{c}}^\dagger P)_i|0\rangle$, $\Theta$ is the projection time, $N_\text{p}$ is the number of particles, and $P$ is a matrix with a dimension of $N\times N_\text{p}$. Usually, we choose the trial wavefunction as the ground state of non-interacting part $H_0=\bm{c}^\dagger K \bm{c}$, and then $P$ is composed of $N_\text{p}$ lowest eigenvectors of $K$. To keep a consistent notation with finite temperature case, we will use $2\Theta$ and $\beta$ interchangeably when necessary. In zero-temperature case, we usually define $B^\rangle(\tau) = B(\tau,0)P$ and $B^\langle(\tau) = P^\dagger B(2\Theta,\tau)$. 
In fast update of zero-temperature DQMC, we keep track of $B^\langle(\tau)$, $B^\rangle(\tau)$ and $\left(B^\langle(\tau)B^\rangle(\tau)\right)^{-1}$ instead of Green's function. After each local update, $B^\rangle(\tau)$ is changed to $(I+\Delta)B^\rangle(\tau)$, and the formula to calculate the determinant ratio is similar to the finite temperature case. If the update is accepted, we update $B^\rangle(\tau)$ and $\left(B^\langle(\tau)B^\rangle(\tau)\right)^{-1}$. The computation complexity is $\mathcal{O}(\beta NN_\text{p}^2)$. Applying the delay update to the zero-temperature case is more convenient when working with the Green's function rather than with $B^\langle(\tau)$, $B^\rangle(\tau)$, and $\left(B^\langle(\tau)B^\rangle(\tau)\right)^{-1}$. In zero-temperature DQMC, we calculate 
 Green's function with
\begin{equation}
\label{eq:0tgr}
G(\tau,\tau) = I - B^\rangle(\tau)\left(B^\langle(\tau) B^\rangle(\tau)\right)^{-1}B^\langle(\tau).
\end{equation}
Based on Green's function, the delay update formula for zero-temperature case is quite similar to finite-temperature case, but also has a little difference. At the beginning of local update in each time slice, we have to calculate the Green's function using Eq.~\eqref{eq:0tgr}. Note that computational complexity of delay update based on Green's function is $\mathcal{O}(\beta N^3)$, which is greater than $\mathcal{O}(\beta N N_\text{p}^2)$. To possibly gain better performance than fast update for zero-temperature case, we require $N_\text{p}$ to be comparable to $N$, which is usually the case we considered in strongly correlated electron lattice models.

At the end of this section, we briefly mention the main flow chart of the DQMC method, as shown below.
\rule{\linewidth}{0.4pt}

\label{tab:flowchart}
Initialization from $\tau=0$ to $\tau=\beta$

Calculate $G(\beta,\beta)$

$\ \ \ \ $do isweep = 1, nsweep
    
$\ \ \ \ \ \ \ \ $Sweep from $\tau=\beta$ to $\tau=0$

$\ \ \ \ \ \ \ \ \ \ \ \ $\textit{Update} at $\tau$

$\ \ \ \ \ \ \ \ \ \ \ \ $\textit{Measurement} at $\tau$

$\ \ \ \ \ \ \ \ \ \ \ \ $\textit{Propagating} from $\tau$ to $\tau'=\tau -a_\tau$

$\ \ \ \ \ \ \ \ \ \ \ \ $\textit{Stabilization} if $\tau'=n\tau_s$

$\ \ \ \ \ \ \ \ $Sweep from $\tau = 0$ to $\tau= \beta$

$\ \ \ \ \ \ \ \ \ \ \ \ $\textit{Update} at $\tau$

$\ \ \ \ \ \ \ \ \ \ \ \ $\textit{Measurement} at $\tau$

$\ \ \ \ \ \ \ \ \ \ \ \ $\textit{Propagating} from $\tau$ to $\tau' = \tau + a_\tau$

$\ \ \ \ \ \ \ \ \ \ \ \ $\textit{Stabilization} if $\tau'=n\tau_s$
  
\rule{\linewidth}{0.4pt}

\noindent where nsweep is the total number of sweeps, and \textit{Stabilization} is performed every $\tau_s$ time slices. In the DQMC, each sweep is made up of four parts: \textit{Update} (to perform sampling of the auxiliary fields), \textit{Measurement} (to measure observables using Green's function), \textit{Propagating} (to evolve the Green's function to the next time slice), and \textit{Stabilization} (to reduce numerical errors caused by successive matrix operations). The details of those parts can be found in Appendix~\ref{sec:app1}. We note that the complexity of the \textit{Measurement} part depends on the physical quantities one wants to calculate. If only equal-time measurements are considered, this part typically exhibits lower complexity; therefore, we omit it from our complexity analysis. The complexities of the other parts are shown in Table~\ref{tab:complexity}, and an extensive discussion on complexity can be found in Appendix~\ref{sec:app2}.

\begin{table}[t!]
\centering
\caption{Computation complexity of DQMC. Each sweep in DQMC is mainly made up of four parts: \textit{Update}, \textit{Measurement}, \textit{Propagating}, and \textit{Stabilization}. The table shows the computational complexity for \textit{Update}, \textit{Propagating}, and \textit{Stabilization}, while the complexity of \textit{Measurement} depends on the physical quantities one wishes to calculate. If we only consider equal-time measurement, this part typically has a smaller complexity, so we omit this part in our analysis. In the table, we list the computational complexities of three different schemes for \textit{Propagating}: FFT (Fast Fourier Transform), Trotter (Trotter decomposition), and ZGEMM (Matrix product).}
\label{tab:complexity}
\begin{tabular}{|c|c|c|c|c|c|}
\hline 
\multirow{2}{*}{DQMC} & \multirow{2}{*}{\textit{Update}} & \multicolumn{3}{c|}{\textit{Propagating}} & \multirow{2}{*}{\textit{Stabilization}}\tabularnewline
\cline{3-5} \cline{4-5} \cline{5-5} 
 &  & FFT & Trotter & ZGEMM & \tabularnewline
\hline
\hline 
finite-T & $\beta N^{3}$ & $\beta N^{2}\ln N$ & $\beta N^{2}$ & $\beta N^{3}$ & $\beta N^{3}$\tabularnewline
\hline 
zero-T & $\beta NN_{\text{p}}^{2}$ & $\beta NN_{\text{p}}\ln N$ & $\beta NN_{\text{p}}$ & $\beta N^{2}N_{\text{p}}$ & $\beta NN_{\text{p}}^{2}$\tabularnewline
\hline 
\end{tabular}
\end{table}


\begin{figure}[b]
	\includegraphics[width=1.0\columnwidth]{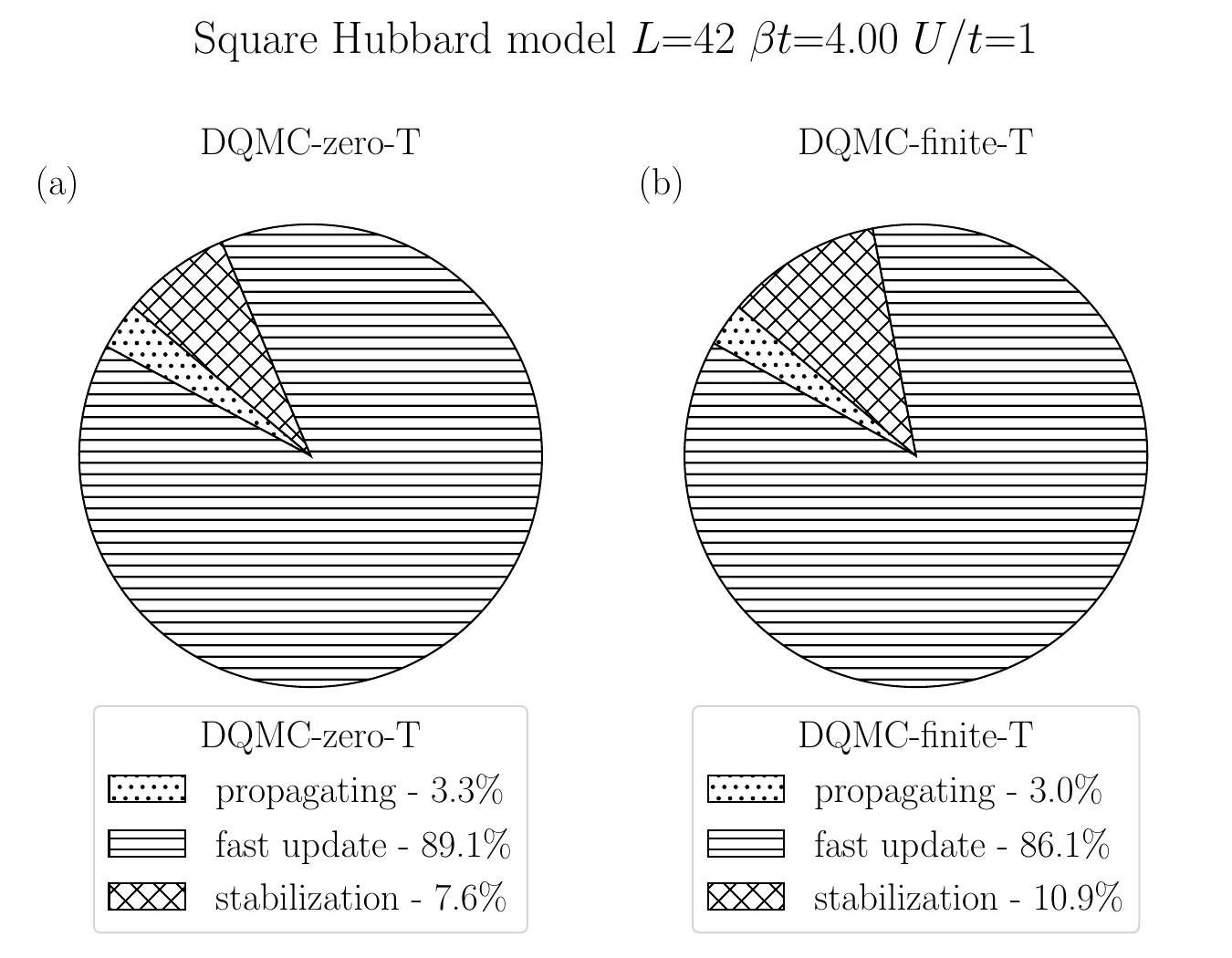}
		\caption{The proportions of total computation time occupied by each process in a DQMC calculation of the Hubbard model on a square lattice. The figure shows computation time for a system size of $42 \times 42$ with model parameters $\beta t=4$ and $U/t=1$. (a) The proportions of each process in DQMC-zero-T. (b) The proportions of each process in DQMC-finite-T. The partition marked by dots represents the propagating part, the partition marked by horizontal lines represents fast updates, and the partition marked by a grid represents the stabilization part.}
	\label{fig:fig2}
\end{figure}

\section{Model and result}
\label{sec:results}
To demonstrate the speedup of the delay update over the fast update in DQMC, let's first consider a scenario where there is only one non-zero diagonal element in $\Delta$ ($k=1$). Applying the delay update to the Hubbard model on a square lattice, as described in~\cite{Gutzwiller1963effect,Hubbard1963electron,arovas2022hubbard,qin2022hubbard}, we have
\begin{align}
H_{tU}=-t\sum_{\langle i,j\rangle,\alpha}(c_{i,\alpha}^{\dagger}c_{j,\alpha}+\text{H.c.})+\frac{U}{2}\sum_i(n_i-1)^2
\end{align}
where $c_{i,\alpha}^{\dagger}$ creates an electron on site $i$ with spin polarization $\alpha=\uparrow/\downarrow$, $n_i=\sum_\alpha c_{i\alpha}^\dagger c_{i\alpha}$ is the fermion number density operator, $t$ is the NN hopping term, and $U$ is the amplitude of onsite Hubbard repulsive interaction. We set $t=1$ as unit of energy and focus on half-filling case, because it does not have sign problem when appropriate HS transformation is used~\cite{Hirsch1983HStransform}. Also note that after HS transformation, spin up and down are block diagonalized, such that spin up and down Green's function can be calculated separately. That is why the Hubbard model is an example of the $k=1$ case in terms of local update. When $U/t=0$, the model at half-filling has a diamond shape of Fermi surface. Turning on $U/t$ induces a metal-insulator transition to an antiferromagnetic insulator phase.

\begin{figure}[tp!]
	\includegraphics[width=1.0\columnwidth]{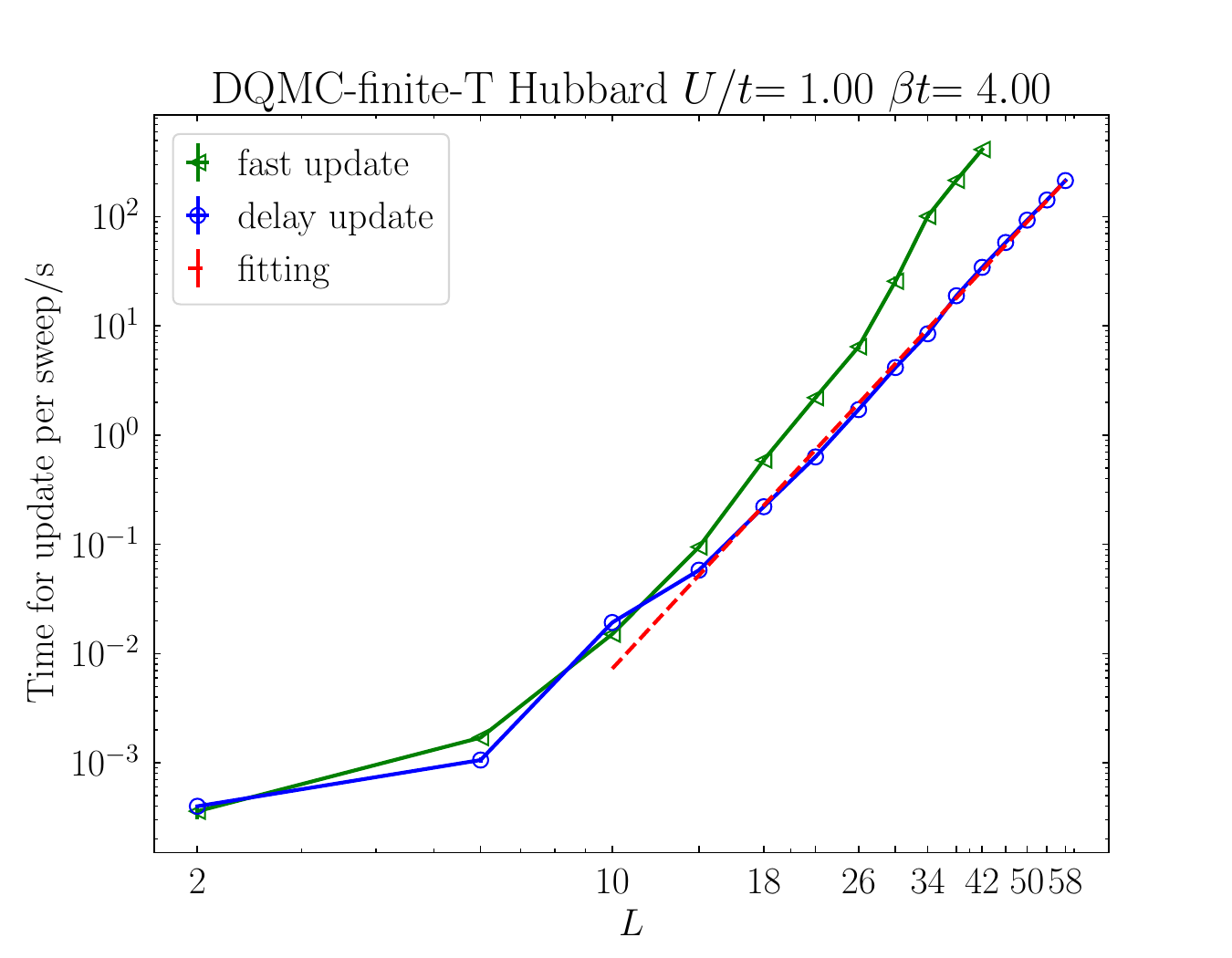}
	\caption{Comparison of the average time for updates per sweep by local updates in the DQMC-finite-T simulation of the Hubbard model, between fast and delay updates. The green line shows the time cost with fast update, while the blue line shows the time cost with delay update. As the system size increases, the acceleration ratio of delay update relative to fast update becomes larger. The red dashed line denotes a rough fitting of the computation complexity of delay update. It is around 6 as expected. If we perform a similar fitting for the blue line, the slope is slightly larger but still around 6. We attribute the slight deviation from 6 to finite size effects. This is also observed in Figs.~\ref{fig:fig4}, \ref{fig:fig6} and~\ref{fig:fig7}.}
	\label{fig:fig3}
\end{figure}	

Before testing the delay update, we first demonstrate the importance of optimizing the fast update component. We estimate the proportions of computation time for various processes in a complete DQMC calculation. To control variables uniformly across all simulations, we set the Hubbard repulsive interaction $U/t=1$, the inverse temperature $\beta t=4$, and the time slice for Trotter decomposition $a_\tau t=0.1$. For each simulation, we run 16 Markov chains, where each Markov chain contains 100 sweeps (divided into 5 bins). To further improve the estimation of computation time, we repeat the simulation 5 times and calculate the average time. This setup applies to all subsequent estimations of computation time in the main text. All simulations were conducted on the Siyuan-1 cluster at Shanghai Jiao Tong University, which we refer to as CPU1 in this work; the configuration of this computer can be found in the Appendix~\ref{sec:app2}. For a system size of $42 \times 42$, as shown in Fig.~\ref{fig:fig2}, the proportion of time taken by fast update exceeds 85$\%$ of the total computation time, regardless of whether it is DQMC-finite-T or DQMC-zero-T. This underscores that optimizing fast updates is essential for accelerating DQMC.

\begin{figure}[tp!]
	\includegraphics[width=1.0\columnwidth]{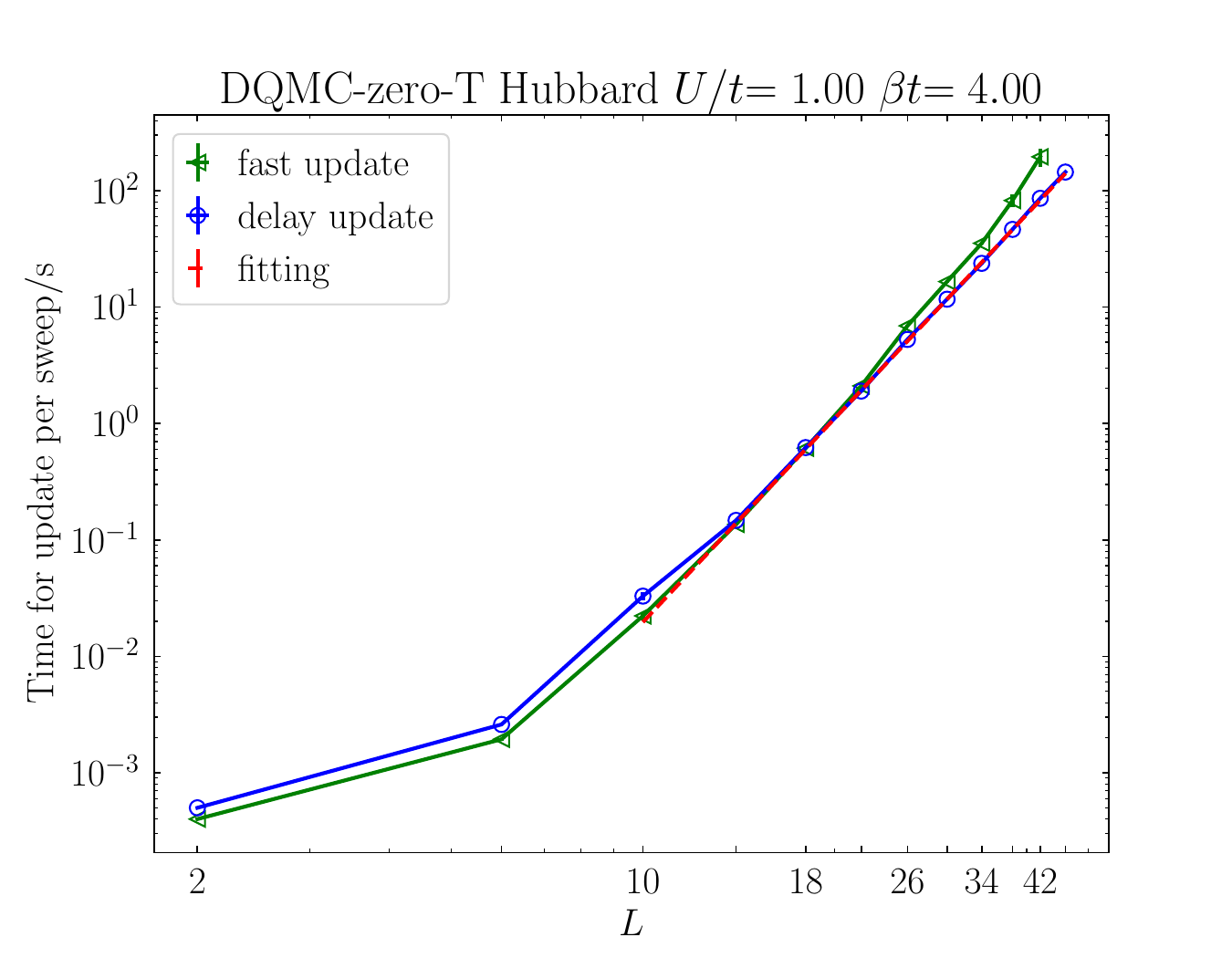}
	\caption{Comparison of the average time for updates per sweep by local updates in DQMC-zero-T simulations of the Hubbard model, comparing fast update and delay update. The green line shows the time cost with fast update, while the blue line shows the time cost with delay update. As the system size increases, the computation time for delay update becomes less than that of fast update.}
	\label{fig:fig4}
\end{figure}

Upon realizing this, we proceed to compare the efficiencies of fast and delay updates. We first consider a finite-T calculation.  We perform DQMC calculations on the Hubbard model with completely identical parameters on systems of different sizes, and we record the average computation time of \textit{Update} per sweep for each system size using different local update algorithms for comparison. A substantial improvement is shown in Fig.~\ref{fig:fig3}. For the same system size, the computation time of delay update is less than that of fast update. Moreover, with an increase in system size, the acceleration factor of delay update continues to grow. When the system size reaches $42 \times 42$, an acceleration factor exceeding tenfold has been achieved. Considering that the proportion of time taken by fast update to the total computation time has exceeded $85\%$, the overall speedup is about sevenfold.
This allows us to elevate the computable system size from  $42 \times 42$ to  $58 \times 58$ while keeping the time cost nearly the same.

Following this, under the same conditions as mentioned above, we compare the delay update with the fast update in DQMC-zero-T.  As discussed earlier, the computational complexity of the fast update in DQMC-zero-T is lower $(\mathcal{O}(\beta N N_p^2) < \mathcal{O}(\beta N^3))$, while the delay update in DQMC-zero-T requires calculating Green's functions before and after each update. As a result, the computation time for the delay update in DQMC-zero-T is longer than that in DQMC-finite-T. Consequently, for small system sizes, the computation time of delay update in DQMC-zero-T is longer than that of fast update. However, as the system size gradually increases, the demand for cache access also grows. In this scenario, delay updates optimized for cache access show acceleration, and as the system size expands, the acceleration factor increases. Within a comparable timeframe, delay updates can simulate systems up to $46 \times 46$ in size, whereas fast updates are limited to system sizes of $42 \times 42$, as illustrated in Fig.~\ref{fig:fig4}.

\begin{figure}[tp!]
	\includegraphics[width=1.0\columnwidth]{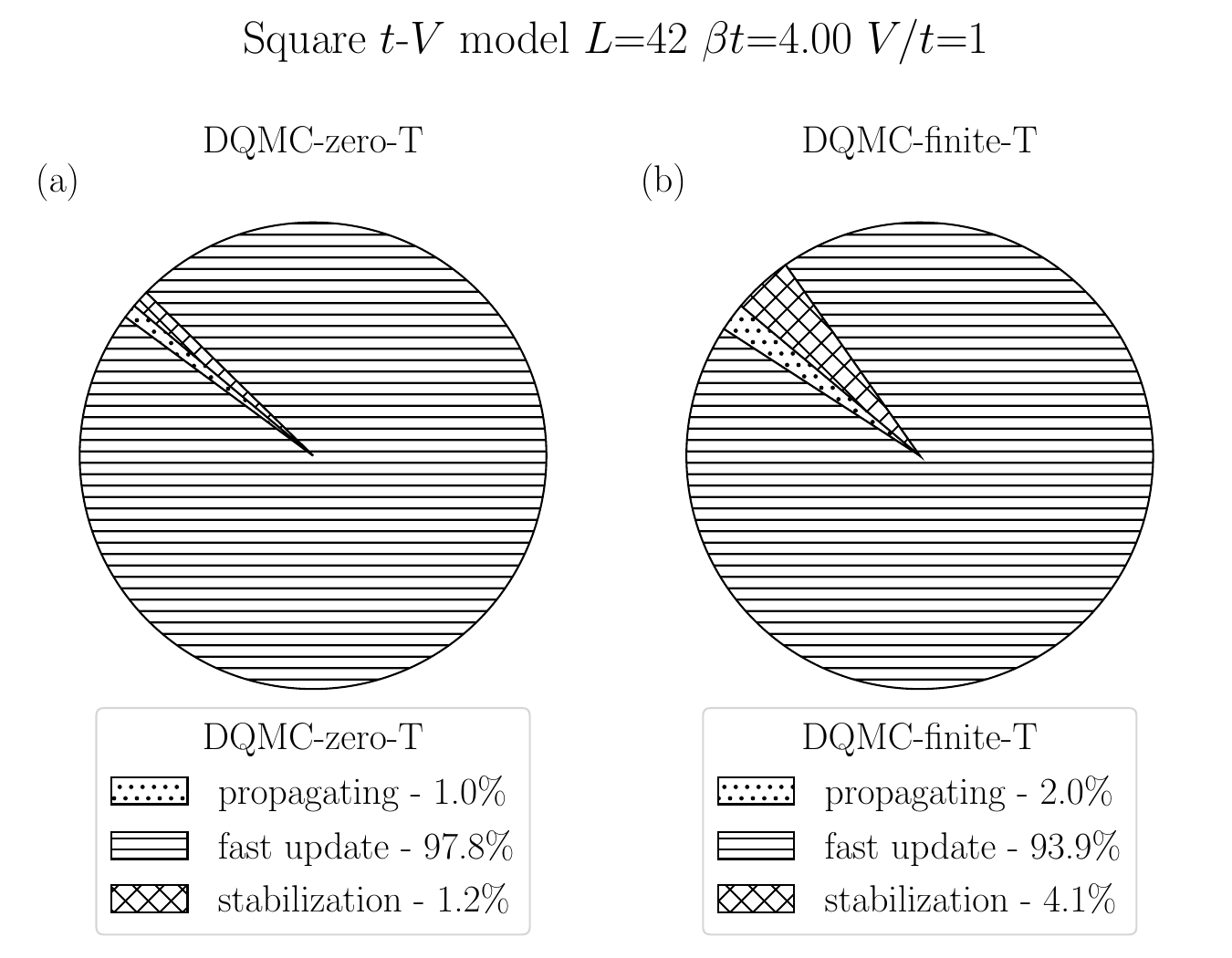}
	\caption{The proportions of total computation time occupied by each process in a DQMC calculation of the spinless $t$-$V$ model on a square lattice. The figure shows computation time for a system size of $42 \times 42$ with model parameters $\beta t=4$ and $V/t=1$. (a) The proportions of each process in DQMC-zero-T. (b) The proportions of each process in DQMC-finite-T. The partition marked by dots represents the propagating part, the partition marked by horizontal lines represents fast updates, and the partition marked by a grid represents the stabilization part.}
	\label{fig:fig5}
\end{figure}

We have demonstrated the superiority of the delay update for onsite interaction. Now, we demonstrate the speedup achieved with extended interaction. We consider a case when $\Delta$ contains only two non-zero diagonal elements $(k=2)$.
Let us consider $t$-$V$ model, which is a spinless fermion model with NN interaction. We consider it on a square lattice and the Hamiltonian is
\begin{align}
H_{tV}=-t\sum_{\langle i,j\rangle}(c_{i}^{\dagger}c_{j}+\text{H.c.})+{V}\sum_{\langle i,j\rangle}(n_i-\frac{1}{2})(n_j-\frac{1}{2})
\end{align}
where $c_{i}^{\dagger}$ creates a fermion on site $i$, $t$ is the nearest-neighbour (NN) hopping term, and $V>0$ is the density repulsive interaction between NN sites. We set $t=1$ as the unit of energy, focus on the half-filling case, and employ a suitable HS transformation to avoid the sign problem\cite{li2015solving,wang2015split}. 
\begin{equation}
\label{eq:tv}
e^{-a_\tau V\left(n_{i}-\frac{1}{2}\right)\left(n_{j}-\frac{1}{2}\right)}=\frac{1}{2}e^{-\frac{a_\tau V}{4}}\sum_{s=\pm1}e^{\alpha_V s(c_{i}^{\dagger}c_{j}+c_{j}^{\dagger}c_{i})}
\end{equation}
 with $\alpha_V=\text{arcosh}(e^{a_\tau V/2})$.
However, this transformation leads to two non-zero off-diagonal elements in $\Delta$ during updates. In order to employ delay updates effectively, we ensure that $\Delta$ only features non-zero elements on its diagonal. To achieve this, we diagonalize $e^{\bm{V}(\bm{s}_{\langle i,j \rangle,l})}$~\footnote{$V(\bm{s}_{\langle i,j \rangle,l})$ is a $2\times 2$ matrix of the coefficient matrix for the interacting part after HS transformation at bond $\langle i,j\rangle$ and time slice $l$.} and propagate the Green's function to a representation where $\Delta$ is diagonal with exactly two non-zero diagonal elements. For the spinless $t$-$V$ model, we set $V/t=1$, keeping all other conditions consistent with the previously discussed Hubbard model. Due to the presence of two non-zero diagonal elements $(k=2)$, the computation time for local updates using fast updates is nearly four times that of the Hubbard model. As a result, the vast majority of computation time is spent on local updates, as shown in Fig.~\ref{fig:fig5}.

\begin{figure}[tp!]
	\includegraphics[width=1.0\columnwidth]{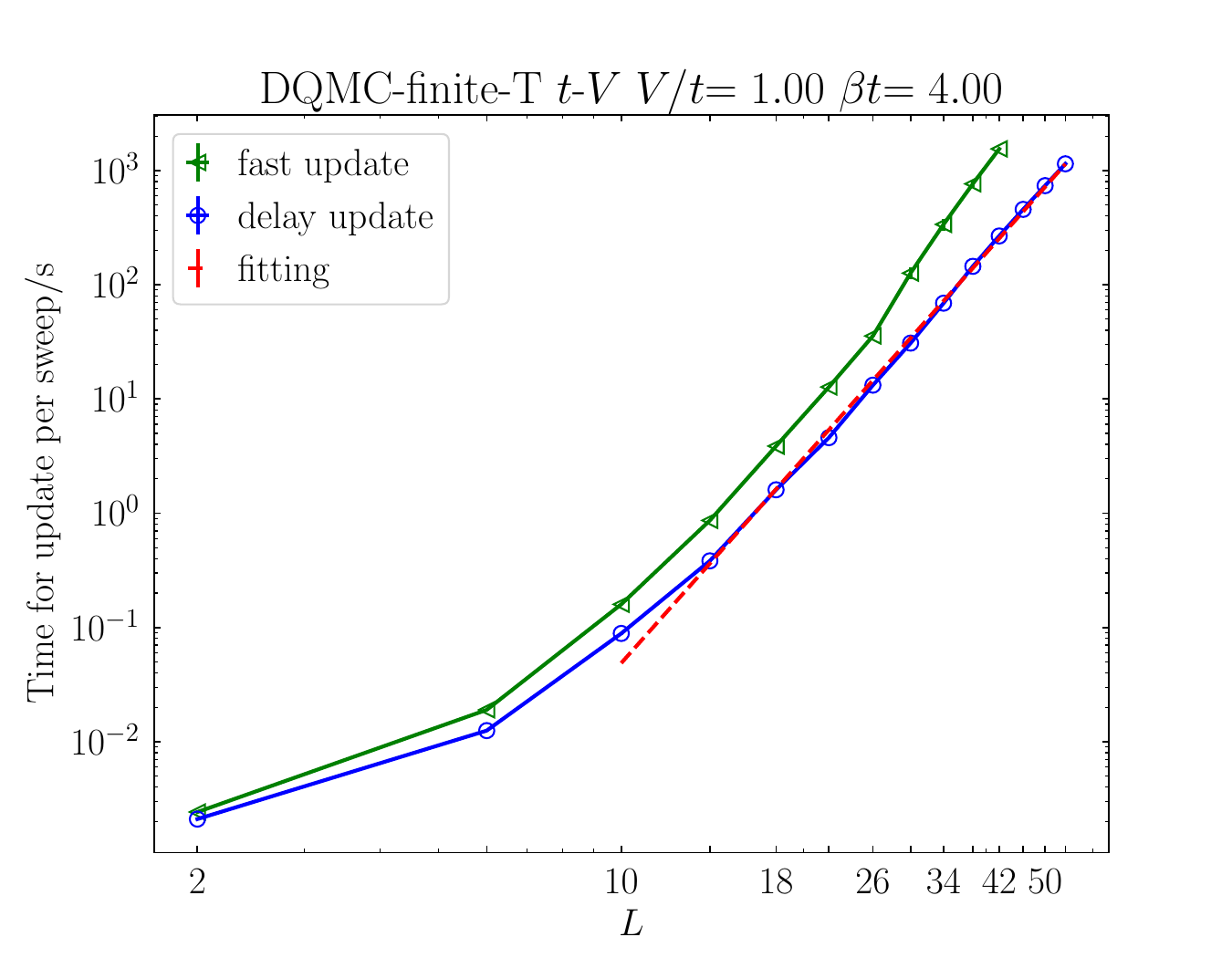}
	\caption{Comparison of the average time for updates per sweep by local updates in DQMC-finite-T simulations of the spinless t-V model, comparing fast update and delay update. The green line shows the time cost with fast update, while the blue line shows the time cost with delay update. The time curve exhibits behavior similar to that observed in the Hubbard model.}
	\label{fig:fig6}
\end{figure}	

We also employed fast update and delay update separately to perform DQMC-finite-T calculations on the spinless $t$-$V$ model. After that, we conducted the same comparison. With an increase in system size, the acceleration factor of delay update continues to grow.  When the system size reaches $42 \times 42$, an acceleration of over sevenfold has been achieved in the \textit{Update} component. This allows us to elevate the computable system size from  $42 \times 42$ to  $54 \times 54$ while keeping the total time cost nearly the same.

\begin{figure}[tp!]
	\includegraphics[width=1.0\columnwidth]{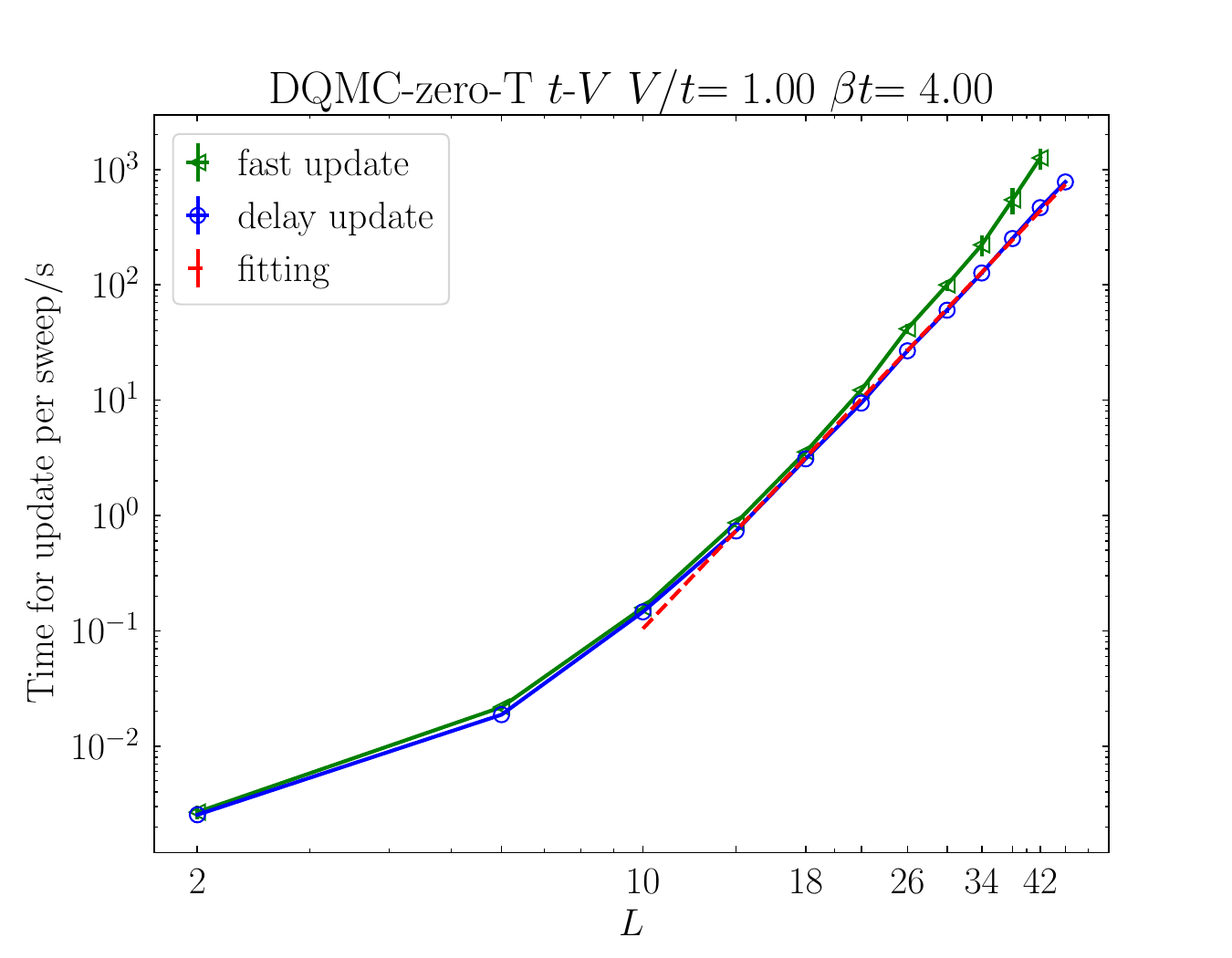}
	\caption{Comparison of the average time for updates per sweep by local updates in DQMC-zero-T simulations of the spinless t-V model, comparing fast update and delay update. The green line shows the time cost with fast update, while the blue line shows the time cost with delay update. Again, the delay update demonstrates superior performance to fast update for larger system sizes.}
	\label{fig:fig7}
\end{figure}	

We tested the acceleration of local updates in DQMC-zero-T using the delay update method in a similar manner. This demonstrates behavior similar to that observed in the Hubbard model under DQMC-zero-T conditions.  Delay update continues to exhibit superior performance compared to fast update; as the system size increases, the acceleration factor of delay update also grows. Furthermore, within nearly the same time frame, the system size for local updates can be increased from $42 \times 42$ (fast update) to $46 \times 46$ (delay update), as shown in Fig.~\ref{fig:fig7}.

\section{Conclusion and discussion}
\label{sec:conclusion}
In this work, we have developed a generalized delay update scheme to accelerate DQMC simulations. It demonstrates significant speedup in updates for both onsite and extended interactions. Based on our tests, the delay update significantly enhances the efficiency of local updates. Moreover, as the system size increases, the computation acceleration factor of delay update continues to grow. By conducting tests on both DQMC-finite-T and DQMC-zero-T, we have found that delay update is particularly well-suited for DQMC-finite-T calculations.  Moving forward, our future work will focus on further optimizing the delay update, especially for DQMC-zero-T. We aim to harness the algorithm's potential within the DQMC-zero-T framework to achieve even better performance. One instructive approach is to consider the sub-matrix algorithm~\cite{nukala2009fast} to reduce the overhead associated with the delay update~\cite{sun2024boosting}. Furthermore, we wish to note that the acceleration provided by the delay update does not depend on whether the model has a sign problem. We believe that the use of delay update can enable DQMC to simulate larger system sizes, thus allowing for improved extrapolation of computational results to the thermodynamic limit and leading to more accurate phase diagrams and phase transitions.

\begin{acknowledgments}
	{\it Acknowledgments}\,---\, X.Y.X. is sponsored by the National Key R\&D Program of China (Grant No. 2022YFA1402702, No. 2021YFA1401400), the National Natural Science Foundation of China (Grants No. 12274289), Shanghai Pujiang Program under Grant No. 21PJ1407200, the Innovation Program for Quantum Science and Technology (under Grant no. 2021ZD0301902), Yangyang Development Fund, and startup funds from SJTU.
\end{acknowledgments}

\appendix

\section{Outline of DQMC}
\label{sec:app1}
In a typical DQMC calculation, a sweep is mainly composed of four parts: \textit{Update},
\textit{Measurement},
\textit{Propagating}, and \textit{Stabilization}. Below, we provide a basic introduction to each component, and the computation complexity of them are summarized in Table~\ref{tab:complexity} of the main text. More details can be found in Refs.~\cite{Blankenbecler1981qmcc,scalapino1981monte,Hirsch1983HStransform,Sugiyama1986,Sorella1988numerical,Sorella1989a,loh1992stable,koonin1997shell,Assaad2008compumanybody}.
\subsection{\textit{Update}}
In QMC simulations, the \textit{Update} component involves the following steps: starting at time slice $\tau$, we attempt to flip the auxiliary fields at this slice one by one. The acceptance ratio is calculated according to the detailed balance condition. If the proposed update is accepted, the entire Green's function is updated in the case of a fast update. For delay update, only the relevant part of the Green's function is updated, and some intermediate vectors are stored. These vectors are used to update the entire Green's function once enough have been accumulated.  The details of fast update and delay update are thoroughly discussed in Sec.~\ref{sec:method}.


\subsection{\textit{Measurement}}
We use Wick’s theorem and the Green’s function to perform the measurement. To avoid confusion, we add the subscript $\bm{s}$ back in this part.
The equal-time Green's function can be used to calculate the energy, order parameters, and various equal-time correlation functions of the system.  The unequal-time Green's functions can be used to calculate dynamic properties of the system. The unequal-time Green's functions are defined as
\begin{align}
G_{\bm{s},ij}(\tau_2,\tau_1)\equiv \langle c_i(\tau_2)c_j^{\dagger}(\tau_1)\rangle_{\bm{s}}
\\
G_{\bm{s},ij}(\tau_1,\tau_2)\equiv -\langle c_j^{\dagger}(\tau_2)c_i(\tau_1)\rangle_{\bm{s}}
\end{align}
where $\tau_2 \ge \tau_1$. We obtain the unequal-time Green's function by applying the time evolution matrix on the equal-time Green's function as follows:
\begin{align}
G_{\bm{s}}(\tau_2,\tau_1)&=B_{\bm{s}}(\tau_2,\tau_1)G_{\bm{s}}(\tau_1,\tau_1) \label{eq:gt2t1}\\
G_{\bm{s}}(\tau_1,\tau_2)&=-B_{\bm{s}}^{-1}(\tau_2,\tau_1)\left[I-G_{\bm{s}}(\tau_2,\tau_2)\right] \label{eq:gt1t2}
\end{align}
Then, we can use Wick's theorem to obtain dynamic correlations of the system. We take the spin correlation as an example. For convenience, we define  a Hermitian conjugation of the equal-time Green's function 
\begin{equation}
G^c_{\bm{s},ij}(\tau,\tau)\equiv \langle c_i^{\dagger}(\tau)c_j(\tau)\rangle_{\bm{s}}
\end{equation}
With the help of spin $SU(2)$ symmetry, the spin correlation function can be written as
\begin{equation}
\begin{aligned}
\langle \mathbf{S}_i\cdot \mathbf{S}_j \rangle_{\bm{s}} &= \frac{3}{2}\langle S^+_i S^-_j\rangle_{\bm{s}} \\
&= \frac{3}{2}\langle c^{\dagger}_{i\uparrow}c_{i\downarrow}c^{\dagger}_{j\downarrow}c_{j\uparrow} \rangle_{\bm{s}}\\
&= \frac{3}{2}\langle c^{\dagger}_{i\uparrow}c_{j\uparrow}\rangle_{\bm{s}} \langle c_{i\downarrow}c^{\dagger}_{j\downarrow} \rangle_{\bm{s}}\\
&=\frac{3}{2}G^c_{\bm{s},ij}G_{\bm{s},ij}
\end{aligned}
\end{equation}
In the same way, we can get the dynamical spin correlation function as $\langle S^+_i(\tau) S^-_j(0)\rangle_{\bm{s}} = -G_{\bm{s},ji}(0,\tau)G_{\bm{s},ij}(\tau,0)$. 

We use Trotter decomposition to calculate the unequal-time Green's functions (Eq.(\ref{eq:gt2t1},\ref{eq:gt1t2})), similar to the case of using Trotter decomposition in the \textit{Propagating} part, the computational complexity is $O(\beta N^2)$ for the DQMC-finite-T and $O(\beta N_p N)$ for the DQMC-zero-T.

\subsection{\textit{Propagating}}
Let's first concentrate on the \textit{Propagating} in DQMC-finite-T. During a sweep in DQMC-finite-T, after completing the \textit{Update} at time slice $\tau_1$, and before moving to the next \textit{Update} at time slice $\tau_2$, where $|\tau_2 -\tau_1|=a_\tau$, we first need to transform the equal-time Green's function from $\tau_1$ to $\tau_2$. This process is called \textit{Propagating}. According to the definition of the equal-time Green's function and the time evolution matrix $B$, we can easily get
\begin{align}
G(\tau_{2},\tau_{2}) & =B(\tau_{2},\tau_{1})G(\tau_{1},\tau_{1})B^{-1}(\tau_{2},\tau_{1})\ \ \text{for}\ \tau_{2}>\tau_{1}\\
G(\tau_{2},\tau_{2}) & =B(\tau_{1},\tau_{2})^{-1}G(\tau_{1},\tau_{1})B(\tau_{1},\tau_{2})\ \ \text{for}\ \tau_{2}<\tau_{1}
\end{align}
As mentioned in the main text, there are three methods for \textit{Propagating}: FFT, Trotter decomposition, and matrix product using ZGEMM from LAPACK (hereafter denoted simply as ZGEMM). In the ZGEMM approach, we perform a matrix product between time evolution $B$ matrices and $G(\tau_1,\tau_1)$, both of them are matrices of $N \times N$, resulting in a computational complexity of $\mathcal{O}(N^3)$ for propagating across one time slice. A whole sweep requires going through $\mathcal{O}(\beta)$ numbers of time slices, therefore, the computational complexity is $\mathcal{O}(\beta N^3)$. If the system has lattice translation symmetry, we can use FFT to reduce the computational complexity. For example, to calculate $e^{-a_\tau K}G(\tau_1,\tau_1)$, we can perform an FFT on the row indices of $G(\tau_1,\tau_1)$, thereby reducing the computational complexity from $\mathcal{O}(N^2)$ to $\mathcal{O}(N\ln N)$ for each column of the resulting matrix.     We need to perform $N$ times of FFT to get all columns of the resulting matrix, so that using FFT to do the propagating for $\mathcal{O}(\beta)$ numbers of time slices reduces the complexity to $\mathcal{O}(\beta N^2\ln N)$. In the end, we consider the \textit{Propagating} using the Trotter decomposition. For example, $B(\tau_2,\tau_1)G(\tau_1,\tau_1)=e^{V(\bm{s})}e^{-a_\tau K}G(\tau_1,\tau_1)$, where $e^{-a_\tau K}G(\tau_1,\tau_1)$ can be Trotter decomposed into $e^{-a_\tau K} \approx \prod_{b}e^{-a_\tau K_{b}}$, where each $e^{-a_\tau K_{b}}$ is a sparse matrix with the non-zero elements $(e^{-a_\tau K_{b}})_{ij}$ ($i$ and $j$ belongs to the two sites of the hopping bond $b$). The computational complexity of the calculation of a certain bond of $e^{-a_\tau K_b}G(\tau_1,\tau_1)$ is $\mathcal{O}(N)$, so the computational complexity of propagating using trotter decomposition involves $\mathcal{O}(\beta N)$ numbers of bonds, leading to the computation complexity of $\mathcal{O}(\beta N^2)$. Similarly, multiplying $e^{V(\bm{s})}$ to the left side of $e^{-a_\tau K_b}G(\tau_1,\tau_1)$ can also be achieved using Trotter decomposition, and it also has the same computational complexity $\mathcal{O}(\beta N^2)$.

Having discussed the \textit{Propagating} in the DQMC-finite-T, now we turn to the DQMC-zero-T. The propagating process in DQMC-zero-T is similar to that in DQMC-finite-T. The only difference between them is the time evolution matrix $B$ applies on $B^\rangle(\tau_1)$ or $B^\langle(\tau_1)$ in DQMC-zero-T, which is a matrix of $N \times N_\text{p}$ or $N_\text{p} \times N$. Taking $B(\tau_2,\tau_1)B^\rangle(\tau_1)$ as an example, the computational complexity for a specific column remains the same as in DQMC-finite-T; however, there are only $N_\text{p}$ columns in $B^\rangle(\tau_1)$, so the computational complexity of the propagating with ZGEMM, FFT and Trotter decomposition is $\mathcal{O}(\beta N^2 N_\text{p})$, $\mathcal{O}(\beta N_\text{p} N \ln N)$ and $\mathcal{O}(\beta N_\text{p} N)$ respectively.

\subsection{Stabilization}
Let's first focus on the \textit{Stabilization} component of DQMC-finite-T. The equal-time Green's function is a key quantity in DQMC-finite-T. It is updated in the \textit{Update} component and propagated in the \textit{Propagating} component; therefore, numerical errors accumulate during these processes. In order to reduce numerical errors, we need to recalculate the equal-time Green's function with the time evolution matrix $B$ at the set intervals. Therefore, the key part is how to obtain the $B$ matrix through stable matrix multiplication. $B$ is a series of matrix multiplications and SVD or QR decomposition is commonly used to stabilize the numerical values of $B$ during the multiplication process~\citep{loh2005numerical,Assaad2008compumanybody}.
For example, we have
\begin{equation}
B(n\tau_s,0)=\mathbf{U}_n
\underbrace{
\left[
\begin{array}{cccc}
\text{\Large X} &  &  &  \\
  & \text{\large X} &  &  \\
  &  & \text{\small X} &  \\
  &  &  & \text{\tiny X} \\
\end{array}
\right]
}_{{\mathbf{D}_n}}
\mathbf{V}_n
\end{equation}
and we can perform singular value decomposition every $\tau_s$ step
\begin{equation}
\begin{aligned}
B((n+1)\tau_s,0)&=B((n+1)\tau_s,n\tau_s)\mathbf{U}_n
\underbrace{
\left[
\begin{array}{cccc}
\text{\Large X} &  &  &  \\
  & \text{\large X} &  &  \\
  &  & \text{\small X} &  \\
  &  &  & \text{\tiny X} \\
\end{array}
\right]
}_{{\mathbf{D}_n}}
\mathbf{V}_n\\
&=\underbrace{
\left[
\begin{array}{cccc}
\text{\Large X} & \text{\large X} & \text{\small X} & \text{\tiny X}  \\
  \text{\Large X} & \text{\large X} & \text{\small X} & \text{\tiny X}  \\
  \text{\Large X} & \text{\large X} & \text{\small X} & \text{\tiny X}  \\
  \text{\Large X} & \text{\large X} & \text{\small X} & \text{\tiny X} \\
\end{array}
\right]
}_{{B((n+1)\tau_s,n\tau_s)\mathbf{U}_n\mathbf{D}_n}}
\mathbf{V}_n\\
&=\mathbf{U}_{n+1}
\underbrace{
\left[
\begin{array}{cccc}
\text{\Large X} &  &  &  \\
  & \text{\large X} &  &  \\
  &  & \text{\small X} &  \\
  &  &  & \text{\tiny X} \\
\end{array}
\right]
}_{{\mathbf{D}_{n+1}}}
\underbrace{
\mathbf{V}'\mathbf{V}_n
}_{{\mathbf{V}_{n+1}}}\\
&=\mathbf{U}_{n+1}\mathbf{D}_{n+1}\mathbf{V}_{n+1}
\end{aligned}
\end{equation}
Correspondingly, for $B(\beta,n\tau_s)$, we also make a similar decomposition, The above $\mathbf{U}_n$, $\mathbf{D}_n$, and $\mathbf{V}_n$ matrices all need to be stored. At imaginary time $\tau=n\tau_s$, we recalculate the equal-time Green's function. From the above decomposition, we have $B(\tau,0)=\mathbf{U}_{R}\mathbf{D}_{R}\mathbf{V}_{R}$ and $B(\beta,\tau)=\mathbf{V}_{L}\mathbf{D}_{L}\mathbf{U}_{L}$, the equal-time Green's function is
\begin{widetext}
\begin{equation}
\begin{aligned}
G\left(n \tau_s, n \tau_s\right)= & {[\mathbf{1}+B(\tau, 0) B(\beta, \tau)]^{-1} } \\
= & {\left[\mathbf{1}+\mathbf{U}_R \mathbf{D}_R \mathbf{V}_R \mathbf{V}_L \mathbf{D}_L \mathbf{U}_L\right]^{-1} } \\
= & \mathbf{U}_L^{-1}\left[\left(\mathbf{U}_L \mathbf{U}_R\right)^{-1}+\mathbf{D}_R\left(\mathbf{V}_R \mathbf{V}_L\right) \mathbf{D}_L\right]^{-1} \mathbf{U}_R^{-1} \\
= & \mathbf{U}_L^{-1}\left[\left(\mathbf{U}_L \mathbf{U}_R\right)^{-1}+\mathbf{D}_R^{\max } \mathbf{D}_R^{\min }\left(\mathbf{V}_R \mathbf{V}_L\right) \mathbf{D}_L^{\min } \mathbf{D}_L^{\max }\right]^{-1} \mathbf{U}_R^{-1} \\
= & \mathbf{U}_L^{-1}\left(\mathbf{D}_L^{\max }\right)^{-1}\left[\left(\mathbf{D}_R^{\max }\right)^{-1}\left(\mathbf{U}_L \mathbf{U}_R\right)^{-1}\left(\mathbf{D}_L^{\max }\right)^{-1}\right. \\
& \left.+\mathbf{D}_R^{\min } \mathbf{V}_R \mathbf{V}_L \mathbf{D}_L^{\min }\right]^{-1}\left(\mathbf{D}_R^{\max }\right)^{-1} \mathbf{U}_R^{-1}
\end{aligned}
\end{equation}
\end{widetext}
where $\mathbf{D}_R^{\min }(i,i)=\min\{\mathbf{D}_R(i,i),1\}$ and $\mathbf{D}_R^{\max }(i,i)=\max\{\mathbf{D}_R(i,i),1\}$. Generally speaking, the selection of the interval for numerical stability $\tau_s$ is related to the severity of numerical instability. It is generally required that the selection of $\tau_s$ should make the numerical error of the equal-time Green's function below $10^{-6}$.

In the end, we discuss the computational complexity of stabilization briefly. At each time when we do the stabilization we need to calculate the equal-time Green's function with $\mathbf{U}_{R}$, $\mathbf{D}_{R}$, $\mathbf{V}_{R}$, $\mathbf{V}_{L}$, $\mathbf{D}_{L}$ and $\mathbf{U}_{L}$, the computational complexity of each stabilization is $\mathcal{O}(N^3)$. We perform stabilization at intervals of $\tau_s$ time slices; therefore, stabilization occurs $\mathcal{O}({\beta}/{\tau_s})$ times during a sweep. Since $\tau_s$ is usually a fixed value, the total computational complexity for stabilization is $\mathcal{O}(\beta N^3)$.

For DQMC-zero-T, the \textit{Stabilization} process is much simpler, as we perform SVD or QR decomposition for $B^{\rangle}(\tau)$ and $B^{\langle}(\tau)$, which are matrices of $N\times N_\text{p}$ or $N_\text{p}\times N$, therefore, the computational complexity for \textit{Stabilization} of this case is $\mathcal{O}(\beta N N_\text{p}^2)$.

\begin{widetext}
\section{The impact of relevant parameters on the delay update}
\label{sec:app2}
Given that the delay update can optimize cache usage, its acceleration varies across different computer hardware configurations. To test how the delay update performs on different platforms, we conducted tests on three computing platforms with x86 architecture CPUs: the Siyuan-1 cluster at Shanghai Jiao Tong University (denoted as CPU1), the $\pi$ 2.0 cluster at Shanghai Jiao Tong University (denoted as CPU2), and a personal desktop computer (denoted as CPU3). The configurations of those platforms are shown in Table~\ref{tab:tab1}.

\begin{table}[h]
  \centering
  \caption{CPU and memory configurations in different platforms.}
  \begin{tabular}{|c|c|}
    \hline
    Computer name & Configurations \\
    \hline
  \multirow{2}{*} {CPU1} & \multicolumn{1}{p{14cm}|}{CPU: 2 $\times$ Intel Xeon ICX Platinum 8358 (2.6GHz, 32 cores, 48MB Intel Cache) \newline Memory: $16 \times 32$ GB TruDDR4 3200 MHz (2R$\times 8$ 1.2V) RDIMM}\\
    \hline
    \multirow{2}{*} {CPU2} & \multicolumn{1}{p{14cm}|}{CPU: 2 $\times$ Intel Xeon Scalable Cascade Lake 6248 (2.5GHz, 20 cores, 27.5MB Intel Cache) \newline Memory: $12 \times 16$ GB Samsung 2666 MHz DDR4 ECC REG }\\
    \hline
    \multirow{2}{*} {CPU3} & \multicolumn{1}{p{14cm}|}{CPU: Intel Core i9-13900K (3.0GHz, 24 cores, 36MB Intel Smart Cache) \newline Memory: $2 \times 32$ GB ADATA XPG Z1 3200 MHz DDR4}\\
    \hline
  \end{tabular}
  \label{tab:tab1}
\end{table}

In the tests, we performed calculations on a square lattice of size $L=30$. For each test, we run 16 Markov chains, where each Markov chain contains 100 sweeps (divided into 5 bins).  We evaluated both DQMC-finite-T and DQMC-zero-T scenarios for the Hubbard model and the spinless t-V model, varying interaction strengths. We recorded the time required per sweep for updates and the acceptance ratio for comparison. The results are summarized in Table~\ref{tab:tab3}.  It is obvious that the delay update has a noticeable acceleration effect on any hardware platform. Interestingly, it seems that acceleration decreases with increasing interaction strength. This illusion arises because, with $a_\tau$ fixed, larger interactions lead to a smaller acceptance ratio, resulting in less frequent use of the delay update.

Next, we test the impact of different temperatures on the acceleration. For simplicity, we used the CPU1 platform to compare the time cost of updates for both the Hubbard model and spinless t-V model at various temperatures in both DQMC-finite-T and DQMC-zero-T scenarios. The results are shown in Table~\ref{tab:tab4}.
It is clear that changes in temperature do not affect the acceleration efficiency of the delay update. This is because an increase in $\beta t$ only increases the number of time slices after Trotter decomposition, while the size of the generated matrices and the time required to update each time slice remain unchanged. As seen, with a tenfold increase in $\beta t$, the time also increases by tenfold both in fast update and delay update scheme, aligning with the computational complexity of local updates being $\mathcal{O}(\beta N^3)$.

\begin{table}[h]
  \centering
  \caption{The average time for updates per sweep by local updates and the acceptance ratio for the Hubbard model and the spinless t-V model with different interaction strengths on different platforms are presented. Both DQMC-finite-T and DQMC-zero-T scenarios are considered. In the table, 'fast' refers to fast update, 'delay' to delay update, and 'ratio' to acceptance ratio.}
  \begin{tabular}{|c|c|c|c|c|c|c|c|}
    \hline
    \multicolumn{8}{|c|}{DQMC-finite-T Hubbard $\beta t=4.00$ $L=30$}\\
    \hline
    computer name & \multicolumn{2}{|c|}{CPU1} & \multicolumn{2}{|c|}{CPU2} & \multicolumn{2}{|c|}{CPU3} &  \\
    \hline
    U/t & delay time(s) & fast time(s) & delay time(s) & fast time(s) & delay time(s) & fast time(s) & ratio(\%)\\
    \hline
    2 & 3.88 & 26.62 & 3.62 & 46.45 & 10.04 & 285.40 & 0.400\\
    \hline
    3 & 3.64 & 22.50 & 3.38 & 44.23 & 9.44 & 261.00 & 0.376\\
    \hline
    4 & 3.46 & 21.67 & 3.23 & 41.41 & 8.40 & 251.80 & 0.356\\
    \hline
    5 & 3.25 & 20.18 & 3.05 & 40.03 & 7.48 & 238.24 & 0.339\\
    \hline
  \end{tabular}
  
  \begin{tabular}{|c|c|c|c|c|c|c|c|}
    \hline
    \multicolumn{8}{|c|}{DQMC-zero-T Hubbard $\beta t=4.00$ $L=30$}\\
    \hline
    computer name & \multicolumn{2}{|c|}{CPU1} & \multicolumn{2}{|c|}{CPU2} & \multicolumn{2}{|c|}{CPU3} &  \\
    \hline
    U/t & delay time(s) & fast time(s) & delay time(s) & fast time(s) & delay time(s) & fast time(s) & ratio(\%)\\
    \hline
    2 & 12.00 & 15.63 & 11.44 & 19.48 & 25.96 & 40.44 & 0.402\\
    \hline
    3 & 11.75 & 16.26 & 11.41 & 19.01 & 25.44 & 43.44 & 0.378\\
    \hline
    4 & 11.56 & 14.93 & 11.49 & 18.55 & 23.72 & 42.84 & 0.357\\
    \hline
    5 & 11.38 & 15.18 & 11.03 & 17.84 & 24.00 & 40.20 & 0.340\\
    \hline
  \end{tabular}
  
  \begin{tabular}{|c|c|c|c|c|c|c|c|}
    \hline
    \multicolumn{8}{|c|}{DQMC-finite-T $t-V$ $\beta t=4.00$ $L=30$}\\
    \hline
    computer name & \multicolumn{2}{|c|}{CPU1} & \multicolumn{2}{|c|}{CPU2} & \multicolumn{2}{|c|}{CPU3} &  \\
    \hline
    V/t & delay time(s) & fast time(s) & delay time(s) & fast time(s) & delay time(s) & fast time(s) & ratio(\%)\\
    \hline
    2 & 25.67 & 128.49 & 24.94 & 158.11 & 75.24 & 759.92 & 0.748\\
    \hline
    3 & 23.79 & 90.63 & 23.14 & 145.69 & 73.40 & 763.95 & 0.691\\
    \hline
    4 & 22.62 & 81.88 & 21.66 & 143.79 & 64.84 & 667.36 & 0.650\\
    \hline
    5 & 21.87 & 83.02 & 21.04 & 122.42 & 58.80 & 522.64 & 0.629\\
    \hline
  \end{tabular}
  
  \begin{tabular}{|c|c|c|c|c|c|c|c|}
    \hline
    \multicolumn{8}{|c|}{DQMC-zero-T $t-V$ $\beta t=4.00$ $L=30$}\\
    \hline
    computer name & \multicolumn{2}{|c|}{CPU1} & \multicolumn{2}{|c|}{CPU2} & \multicolumn{2}{|c|}{CPU3} &  \\
    \hline
    V/t & delay time(s) & fast time(s) & delay time(s) & fast time(s) & delay time(s) & fast time(s) & ratio(\%)\\
    \hline
    2 & 58.35 & 89.05& 56.81 & 94.43 & 136.96 & 196.24 & 0.751\\
    \hline
    3 & 56.21 & 86.03 & 51.98 & 91.26 & 130.24 & 230.24 & 0.693\\
    \hline
    4 & 54.76 & 83.57& 51.09 & 86.28 & 128.44 & 204.64 & 0.650\\
    \hline
    5 & 53.87 & 82.44 & 49.38 & 87.43 & 134.48 & 222.00 & 0.622\\
    \hline
  \end{tabular}
  \label{tab:tab3}
\end{table}

\begin{table}[h]
  \centering
  \caption{The average time for updates per sweep by local updates in  the Hubbard model and spinless t-V model with different $\beta t$ on CPU1.	}
  \begin{tabular}{|c|c|c|c|c|}
    \hline
    \multicolumn{5}{|c|}{DQMC Hubbard $U/t=1.00$ $L=30$ local update time per sweep}\\
    \hline
      & \multicolumn{2}{|c|}{$\beta t=4.00$} & \multicolumn{2}{|c|}{$\beta t=40.00$}  \\
    \hline
      & delay time(s) & fast time(s) & delay time(s) & fast time(s) \\
    \hline
    DQMC-finite-T & 4.16 & 23.96 & 41.29 & 254.20\\
    \hline
    DQMC-zero-T & 11.69 & 16.57 & 117.76 & 164.13\\
    \hline
  \end{tabular}
  
  \begin{tabular}{|c|c|c|c|c|}
    \hline
    \multicolumn{5}{|c|}{DQMC t-V $V/t=1.00$ $L=30$ local update time per sweep}\\
    \hline
      & \multicolumn{2}{|c|}{$\beta t=4.00$} & \multicolumn{2}{|c|}{$\beta t=40.00$}  \\
    \hline
      & delay time(s) & fast time(s) & delay time(s) & fast time(s) \\
    \hline
    DQMC-finite-T & 30.94 & 120.29 & 287.33 & 1203.19\\
    \hline
    DQMC-zero-T & 58.99 & 94.34 & 589.65 & 943.65\\
    \hline
  \end{tabular}
  \label{tab:tab4}
\end{table}
\end{widetext}
\clearpage
\bibliographystyle{apsrev4-2}
\bibliography{main.bib}

\end{document}